\documentclass[9pt,twocolumn,twoside]{opticajnl}
\journal{opticajournal} 

\setboolean{shortarticle}{true}

\usepackage{lineno}

\title{Adaptive sampling strategy for tolerance analysis of freeform optical surfaces based on critical ray aiming}

\author[1]{Rundong FAN}
\author[1]{Shili Wei}
\author[1]{Zhuang Qian}
\author[2]{Huiru Ji}
\author[2]{Hao Tan}
\author[2]{Yan Mo}
\author[1,2,3,*]{Donglin Ma}

\affil[1]{School of Optical and Electronic Information, Huazhong University of Science and Technology, Wuhan, Hubei 430074, China}
\affil[2]{MOE Key Laboratory of Fundamental Physical Quantities Measurement $\&$ Hubei Key Laboratory of Gravitation and Quantum Physics, PGMF and School of Physics, Huazhong University of Science and Technology, Wuhan 430074, China}
\affil[3]{Shenzhen Huazhong University of Science and Technology, Shenzhen 518057, China}

\affil[*]{madonglin@hust.edu.cn}

\begin{abstract}
The tolerance analysis of freeform surfaces plays a crucial role in the development of advanced imaging systems. However, the intricate relationship between surface error and imaging quality poses significant challenges, necessitating dense sampling of featured rays during the computation process to ensure an accurate tolerance for different fields of view (FOVs). Here, we propose an adaptive sampling strategy called "Critical Ray Aiming" for surface tolerance analysis. By identifying the most sensitive ray to wave aberration at each surface point, our methodology facilitates flexible sampling of the FOVs and entrance pupil (EP), achieving computational efficiency without compromising accuracy in determining tolerable surface error. We demonstrate the effectiveness of our method through tolerance analysis of two different freeform imaging systems. 
\end{abstract}

\setboolean{displaycopyright}{false} 

\begin{document}

\maketitle
Over the past decade, freeform optics \cite{bauer2018starting,rolland2021freeform} has revolutionized compact and high-performance optical systems. Freeform surfaces can be defined as surfaces with no axis of rotational invariance (within or beyond the optical part). The development of freeform optical surfaces has grown substantial changes in imaging systems \cite{tan2023freeform,yan2020general,yang2017automated} and non-imaging systems \cite{wei2023sculpting,wu2016direct,bosel2019compact}. Nevertheless, the high degrees of freedom of freeform surfaces also increase their manufacturing difficulty \cite{zhu2018review,li2017machining}, making it challenging to achieve design performance. The tolerance analysis of optical surfaces serves as the bridge between optical design and optical manufacturing, representing a perennial research subject.

A common solution of surface tolerance is adjusting the root-mean-square (RMS) or peak-to-valley (PV) value of entire surfaces by multiple Monte-Carlo experiments \cite{graaff1993condensed,hokr2015modeling,duan2011monte}. This method, grounded in statistical and probabilistic theory, has been deemed the most effective approach for analyzing the tolerances of optical components, particularly suitable for mass-produced optical elements. Recently, Deng et al. \cite{deng2022local} proposed a local tolerance calculation method.  Their research unveiled that optical surface tolerances exhibit local characteristics, with different regions of the optical surface displaying varying tolerance values due to differences in sensitivity to optical system performance. Consequently, this method proves more accurate than global tolerance analysis relying solely on overall RMS or PV values. 
\begin{figure}[ht]
\centering
\fbox{\includegraphics[width=\linewidth]{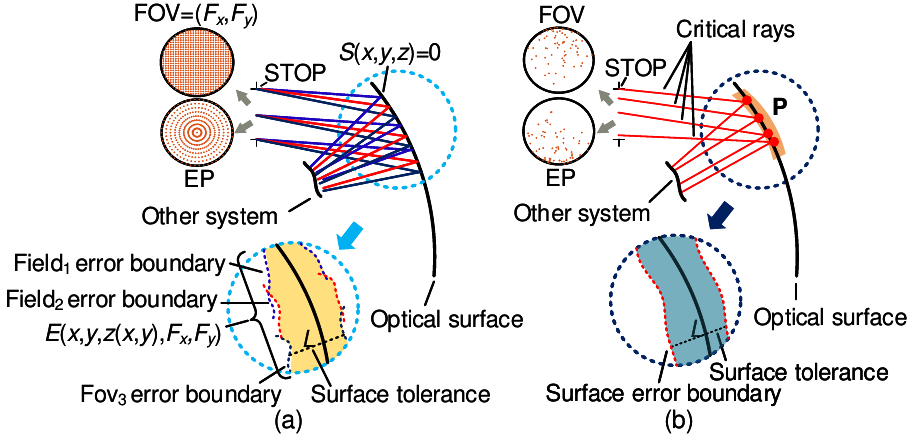}}
\caption{Optical surface tolerance calculation: (a) dense uniform rays sampling; (b) critical rays sampling}
\label{fig:false-color}
\end{figure}

In Fig. 1(a), it is illustrated that for a given surface $S$($x$, $y$, $z$)=0 within an optical system, each field corresponds to a change in surface sag denoted by $E$($x$, $y$, $z$, $F_x$, $F_y$). This change results in the wave aberration of each field reaching a critical value, representing the maximum tolerable wave aberration degradation. To maintain consistent imaging performance across the entire field of view (FOV), the tolerance for the optical surface is determined as the envelope of the tolerances associated with each specific field. The envelope of these surface ranges of each field can be expressed as:

\begin{equation}\begin{cases}E(x,y,z,F_x,F_y)=0\\\partial E/\partial F_x=0,\partial E/\partial F_x=0\end{cases},\end{equation}where $F_x$, $F_y$ are constant, representing the $F^{th}$ field in the $x$ and $y$ directions as $\mathbf{F}$=($F_x$,$F_y$). Different values of $F_x$ and $F_y$ correspond to distinct surfaces.

During the calculation process, the FOV and entrance pupil (EP) should be uniformly and densely sampled to ensure the accuracy of the tolerance solution. However, excessive sampling of rays can notably decrease computational efficiency. In this letter, we propose a critical ray computation method. The critical ray denotes the ray most sensitive to wave aberration at these points $\mathbf{P}$ traversing through the optical surface, corresponding to the critical points of the surface family. By analyzing the critical sampling rays corresponding to the sampling points of the optical surface and calculating the critical tolerance of each optical surface, the surface tolerance can be determined at once, as illustrated in Fig. 1(b). Furthermore, we showcase the effectiveness of our method by conducting tolerance analysis on two distinct freeform imaging systems.

Figure 2(a) depicts a surface slightly perturbed from its nominal position, based on geometric optical perturbation models \cite{rimmer1970analysis}, the correspondence between wave aberration and surface error can be expressed as:
\begin{equation}\delta W_i^F=2Ln\cos^2I_i^F,\end{equation}
Eq. (2) pertains to the reflective surfaces, and while for the refractive surface, the equation $\delta W_i^F=L\cos I_i^F(n\cos I_i^F-n^{\prime}\cos I_i^{\prime F})$ applies, where $n$ and $n^{\prime}$ are the refractive indices on the incident and refracted sides, respectively, $I_{i}^{F}$ and $I_{i}^{\prime F}$ are the angles of incidence and refraction of the $i^{th}$ ray $\mathbf{R_i}$ of the field $F$ at point $\mathbf{P}$. The $L$ is the distance between point $\mathbf{P}$ and point $\mathbf{P^{\prime}}$. Points $\mathbf{P}$ and $\mathbf{P^{\prime}}$ are the intersection points of ray $\mathbf{R_i}$ with the nominal surface and the perturbed surface, respectively.

By sampling the FOV and EP rays, multiple rays pass through at a point $\mathbf{P}$ on the optical surface, as depicted in Fig. 2(b). These rays exhibit different optical path difference (OPD) distributions at the exit pupil, with the ray possessing the largest wave aberration termed the critical ray $\mathbf{R_c}$. Under identical wave aberration boundaries, the $\mathbf{R_c}$ ray corresponds to the minimum surface error $d_{min}$ at point $\mathbf{P}$.

\begin{figure}[ht]
\centering
\fbox{\includegraphics[width=\linewidth]{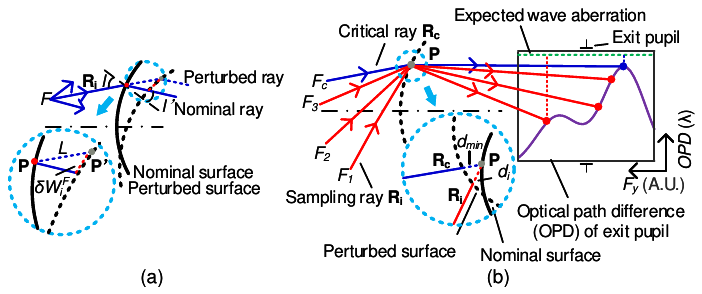}}
\caption{Perturbed optical surface and critical ray: (a) perturbed optical surface; (b) critical ray.}
\label{fig:false-color}
\end{figure}

We define the input 4D rays parameterized by the vector ($F_x$, $F_y$, $P_x$, $P_y$). As depicted in Fig. (3), the $I_{m}^{i}$ and $I_{m}^{\prime i}$ are the angles of incidence and refraction of the $i^{th}$ ray $\mathbf{R_i}$ on the $m^{th}$ surface at point $\mathbf{S_m}$, respectively.

Based on the above considerations, the process of solving critical rays can be reformulated as a numerical problem and can be expressed as:

\begin{equation}
\boldsymbol{\theta^*}=\underset{\boldsymbol{\theta}}{\arg \min } MF(\boldsymbol{\theta}),
\end{equation} 
where $\boldsymbol{\theta}$ is the optimization variable, which represents the parameters ($F_x$, $F_y$, $P_x$, $P_y$). The loss function $MF(\boldsymbol{\theta})$, for each point on the optical surface is segmented into two parts, denoted as $MF(\boldsymbol{\theta})=\omega_1L(\boldsymbol{\theta})+\omega_2C(\boldsymbol{\theta})$. The first part is employed to compute the critical ray, which is the most sensitive to wave aberration, while the second part is utilized to aim the ray, ensuring that the optimized ray is positioned at a fixed surface point. The weights assigned to these two parts are denoted as $\omega_{1}$ and $\omega_{2}$, respectively. The first part of $MF(\boldsymbol{\theta})$ can be obtained by Eq. (2), and it can be expressed as $L(\boldsymbol{\theta})=\left|(W_i-W_{exp})/2n\cos^2I_i\right|$, where $|\cdot|$ is modulo operations, and $W_i$ and $W_{exp}$ are the wave aberration of $i^{th}$ ray and expected wave aberration, respectively.

Traditional ray aiming is an algorithm designed to solution rays aim the STOP of an optical system. It first aims the optical EP and adjusts the coordinates or the direction cosine of the incident ray vector via ray tracing and iterative methods, ensuring that the ray through the correct position of STOP. This method finds widespread application in current design software, such as Zemax. Building upon the principles of ray aiming, we propose a critical ray aiming algorithm based on numerical iteration. Specifically, by aiming the optical surface points $\mathbf{S_m}$, we iteratively refine the parameters ($F_x$, $F_y$, $P_x$, $P_y$) until we determine the critical ray,($F_x^{critical}$, $F_y^{critical}$, $P_x^{critical}$, $P_y^{critical}$), that passes through the designated surface point $\mathbf{S_m}$.

\begin{figure}[ht]
\centering
\fbox{\includegraphics[width=\linewidth]{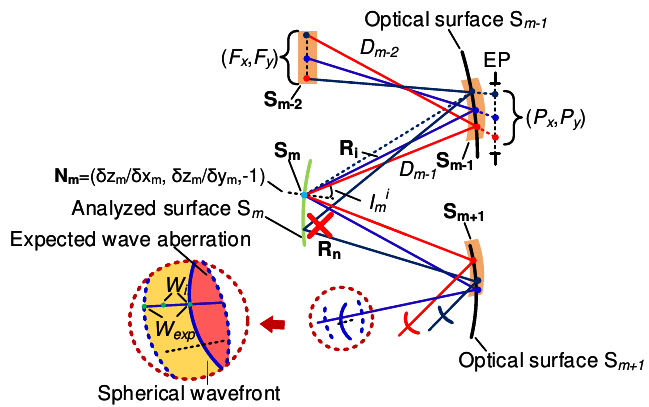}}
\caption{Theoretical diagram of the critical ray evaluation and ray aiming}
\label{fig:false-color}
\end{figure}

Before commencing the calculation, we employ fitting functions $f_x$ and $f_y$ to approximate the $x$ and $y$ coordinates of all optical surfaces within the optical system with 4D ray ($F_x$, $F_y$, $P_x$, $P_y$). As illustrated in Fig. 3, the larger incidence angle of the connection line between point $\mathbf{S_m}$ and point $\mathbf{S_{m-1}}$ leads to a smaller value of the first part of the loss function $MF(\boldsymbol{\theta})$. However, since ray $\mathbf{R_n}$ fails to intersect with point $\mathbf{S_m}$, it is deemed invalid. Hence, a ray aiming constraint must be consistently applied throughout the numerical iteration process to ensure that the ray is accurately directed towards the surface point $\mathbf{S_m}$. Fermat’s principle \cite{duerr2021freeform} is a necessary condition for the existence of a ray. Therefore, we utilize Fermat’s principle on the $(m-1)^{th}$ surface to constrain the solution of the critical ray in the computation process. The equation can be expressed as:

\begin{equation}\begin{cases}\partial D_x=\partial(OPL_{m-1}(D_{m-2},D_{m-1}))/\partial x=0\\\partial D_y=\partial(OPL_{m-1}(D_{m-2},D_{m-1}))/\partial y=0\end{cases},\end{equation}
where optical path length (OPL) between $D_{m-2}$ and $D_{m-1}$ can be obtained using the equation $OPL_{m-1}(D_{m-2},D_{m-1})=n_{m-2}D_{m-2}+n_{m-1}D_{m-1}$. And the distances $D_{m-2}$ and $D_{m-1}$ can be determined using the three points $\mathbf{S_{m-2}}$, $\mathbf{S_{m-1}}$, and $\mathbf{S_m}$ by $D_{m-2}(\mathbf{S}_{m-2},\mathbf{S}_{m-1})=\left\|\mathbf{S}_{m-2}-\mathbf{S}_{m-1}\right\|_{2}$ and $D_{m-1}(\mathbf{S}_{m-1},\mathbf{S}_{m})=\left\|\mathbf{S}_{m-1}-\mathbf{S}_{m}\right\|_{2}$, where $\|\cdot\|_{2}$ represents the second vector norm operation, calculated using the distance equation between two points. And $n_{m-2}$, $n_{m-1}$ are the refractive index of the medium before and after the $(m-1)^{th}$ optical surface.

\begin{figure}[ht]
\centering
\fbox{\includegraphics[width=\linewidth]{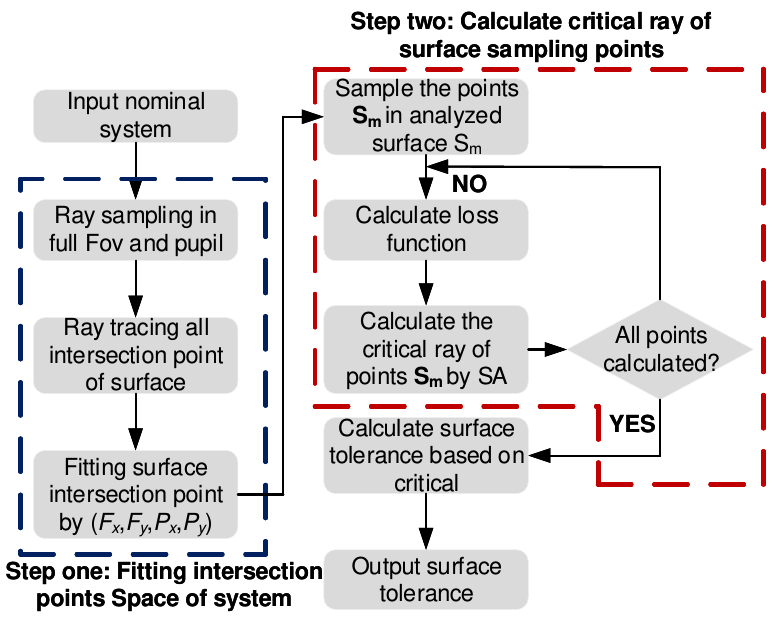}}
\caption{Flowchart of the computation process}
\label{fig:false-color}
\end{figure}

Therefore, the second part of loss function $MF(\boldsymbol{\theta})$ in Eq. (3) can be obtained by Eq. (4), and it can be expressed as $C=\sqrt{(\partial D_{x}^{i})^{2}+(\partial D_{y}^{i})^{2}}$. Above process can be implemented using the simulated annealing (SA) global method in Matlab.

The proposed method calculates the critical ray corresponding to the surface sampling points Sm before the surface tolerance calculation. Once the critical light of each intersection point is determined, it can be further traced and analyzed using other tolerance analysis models, such as point-by-point calculation \cite{deng2022local}. Here, the surface tolerance is calculated rapidly using the tolerance calculation method proposed in our previous work \cite{fan2024surface}. It is noted that critical rays should be determined during each outer loop, as proposed in our previous work. Based on the above considerations, the determination of the critical ray involves a two-step process. First, the coordinate space of the system is fitted using ($F_x$, $F_y$, $P_x$, $P_y$).  Second, the surface is sampled, and the critical ray corresponding to each sampling point is calculated. The flowchart depicting the obtaining process of surface tolerance based on a critical ray is presented in Fig. 4 

\begin{figure}[ht]
\centering
\fbox{\includegraphics[width=\linewidth]{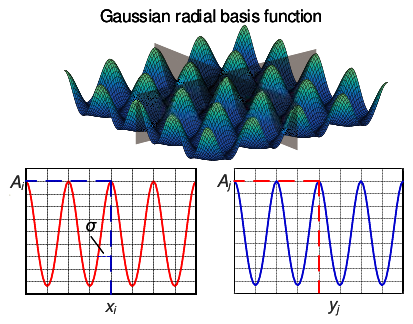}}
\caption{Diagram of gaussian radial basis function}
\label{fig:false-color}
\end{figure}

In this letter, the intersection space between the optical surface and the wavefront is approximated using Gaussian radial basis functions \cite{cakmakci2008optimal}. It should be noted that the fitting process of Gaussian radial basis functions can be accelerated by using a sparse matrix algorithm. 

\begin{equation}f=\sum_{j=1}^{T}\sum_{k=1}^{T}\sum_{l=1}^{T}\sum_{r=1}^{T}A_{j,k,l,j}e^{\frac{-((F_{x}-F_{x}^{j})^{2}+(F_{y}-F_{y}^{k})^{2}+(P_{x}-P_{x}^{l})^{2}+(P_{y}-P_{y}^{r})^{2})}{2\sigma^{2}}},\end{equation}
where $F_{x}^{j}$, $F_{y}^{k}$, $P_{x}^{l}$, and $P_{y}^{r}$ are the Gaussian kernel sampling centers of ($F_x$, $F_y$, $P_x$, $P_y$) respectively. $T$ represents the number of Gaussian kernel sampling center items. The ${\sigma}$ and $A$ are the width and peak value of the Gaussian function basis, which is showcased in Fig. 5. It is worth noting that the fitting function in the algorithm described herein is also applicable to other functions, such as Laguerre-Gaussian basis functions, $XY$ polynomials, and so on.

\begin{figure}[ht]
\centering
\fbox{\includegraphics[width=\linewidth]{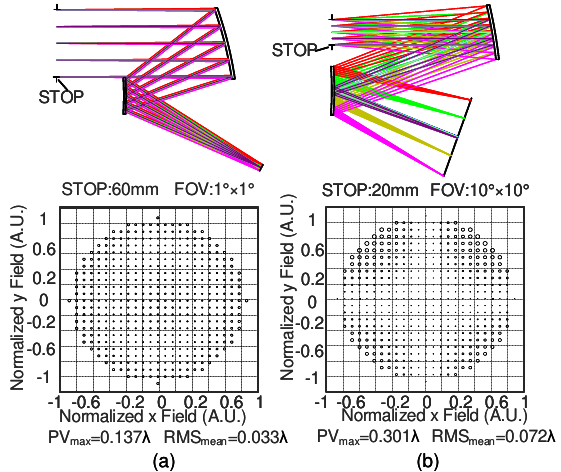}}
\caption{Two-mirror off-axis system layout and field map of the RMS wave aberration; (a) system 1; (b) system 2}
\label{fig:false-color}
\end{figure}

\begin{figure*}[ht]
\centering
\fbox{\includegraphics[width=\linewidth]{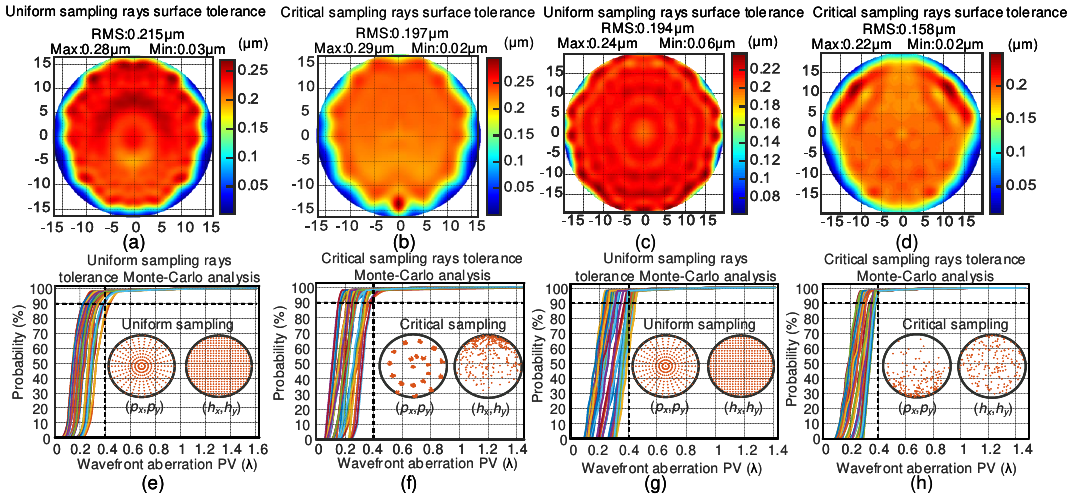}}
\caption{The results of surface sag tolerance and forward Monte Carlo analysis of the second mirror in two systems are depicted as follows: (a) sampling ray tolerance spectrum of System 1; (b) critical ray tolerance spectrum of System 1; (c) sampling ray tolerance spectrum of System 2; (d) critical ray tolerance spectrum of System 2; (e) Monte-Carlo analysis for sampling ray tolerance in System 1; (f) Monte-Carlo analysis for critical ray tolerance in System 1; (g) Monte-Carlo analysis for sampling ray tolerance in System 2; (h) Monte-Carlo analysis for critical ray tolerance in System 2}
\label{fig:false-color}
\end{figure*}

To demonstrate the effectiveness of the proposed method, we analyzed the surface tolerance of second freeform two-mirror systems operating in the infrared (1 \textmu m) with a focal length of 400mm. They have aperture diameters of 60mm and 20mm, and FOVs of 1°×1°and 10°×10°, respectively. The systems are symmetrical about the $YOZ$ plane, with the STOP located on the first surface. All mirrors are freeform surfaces described by $XY$ polynomials up to the fifth order. The system layouts and the RMS wave aberration field maps for the two systems are shown in Figs. 6(a) and (b), respectively. The FOVs were uniformly sampled using a rectangular grid, resulting in a total of 100 field samples for each system. Radial sampling of ray data was performed for each field, resulting in a total of 70 samples for each field. Therefore, the sampling rays for the tolerance computation analyzed using the intensive sampling method are 100×70. Here, we use the coordinates of the intersection points obtained from the densely sampled ray trace for a 4D polynomial fit. It is noted that we can also fit the polynomial with more sampled data. Then, we mesh the second mirrors of the two optical systems to calculate the critical ray, using a sampling quantity of 40×40.

The maximum wave aberration PV value of the two systems is 0.301$\lambda$. To achieve acceptable imaging performance for full FOV, a wave aberration PV threshold of 0.4$\lambda$ is employed during operation. Then, to analyze the error in the wider frequency band, the surface error was represented using Gaussian radial basis functions ~\cite{cakmakci2008optimal}, with 1500 terms of Gaussian centers sampled and ${\sigma}$ of 1. Surface tolerances for the second mirror of both optical systems are computed using the tolerance calculation method ~\cite{fan2024surface} for uniformly and critically sampled rays, as illustrated in Figs. 7(a)$\sim$(b) and (c)$\sim$(d). We found that the tolerance calculated using critical sampling rays exhibits a smaller RMS value than that of the uniformly sampled rays, indicating a tighter sag tolerance of the optical surface. The surface sag tolerance of the sampling rays in the two systems is 0.215\textmu m and 0.194\textmu m, respectively, while the tolerance of critical rays in the two systems is 0.197\textmu m and 0.158\textmu m. Additionally, while the computational steps increase with critical ray sampling compared to dense ray sampling, the total time required to solve local tolerances using critical ray sampling is less due to the significantly reduced number of sampling rays used for calculating surface tolerances, as illustrated in Fig. 8.

To verify the calculation accuracy of critical ray surface tolerances, forward Monte Carlo experiments were conducted on the two systems. Gaussian radial basis functions were employed to generate the surface error of the secondary mirror, with parameters consistent with those used in computing the surface tolerance. We randomly selected 2000 surface errors and individually applied them to the second mirror. Subsequently, we observed the PV value of wave aberration across the two systems. In each Monte Carlo test group, we analyzed the PV value of wave aberration for 60 sampling fields. Figures 7(e)$\sim$(h) illustrate the Monte Carlo results for the two systems. In these results, the wave aberration PV value for the critical rays remains below the preset threshold in 92$\%$ of the experimental groups across the entire FOV. However, in the Monte Carlo experimental group of uniformly and densely sampled rays corresponding to surface tolerances, some fields will exceed the PV threshold of wave aberration. Therefore, the local tolerances of optical surfaces can be calculated more accurately using critical sampling rays.

\begin{figure}[ht]
\centering
\fbox{\includegraphics[width=\linewidth]{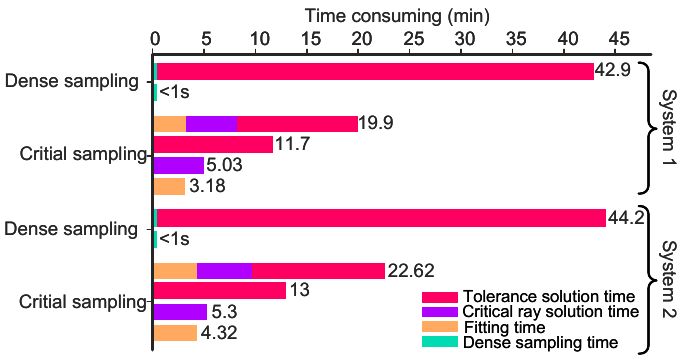}}
\caption{Diagram of the time-consuming picture with two sampling strategies}
\label{fig:false-color}
\end{figure}

As shown in Figs. 7(a)$\sim$(d), the difference in local tolerance RMS calculated by the two rays sampling methods in system 1 is less pronounced than in system 2. To further analyze the influence of two ray sampling methods on the tolerances of optical systems with large FOVs and large apertures, we partitioned the maximum aperture of System 1 into proportions of [0.2, 0.5, 0.7, 1]. We accordingly divided the FOV of System 2 and subsequently observed the differences in tolerance calculation between the primary and secondary mirrors in both systems. As shown in Fig. 9, the system fields exhibit higher sensitivity to tolerance calculation error, evidenced by a steep curve. Conversely, the system aperture demonstrates lower sensitivity to such discrepancies, manifesting as a flatter curve. Hence, it is more necessary to use critical ray sampling for the calculation of surface local tolerances with large-field optical systems.

\begin{figure}[ht]
\centering
\fbox{\includegraphics[width=\linewidth]{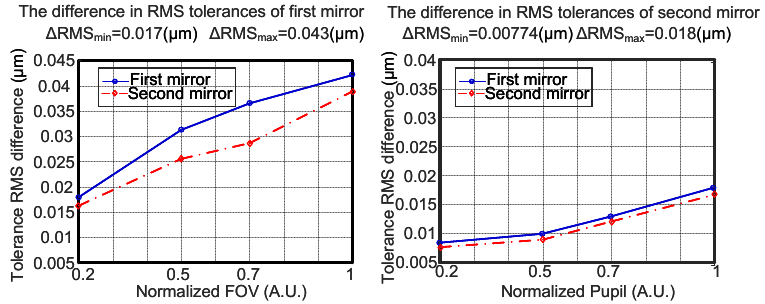}}
\caption{Differences in surface sag tolerance between the two mirrors: (a) system 1; (b) system 2}
\label{fig:false-color}
\end{figure}

We introduced a critical ray calculation method that facilitates the reverse computation of optical surface tolerances for multiple rays intersecting simultaneously. By integrating wave aberration analysis, freeform surface design, and ray tracing with 4D ray modeling and mathematical numerical iterative computation, we have developed an effective and promising approach to determining optical surface sag tolerances. Two freeform surface reflective system sag tolerance calculation examples were analyzed. In a single calculation process, the critical ray solution is solved for a single surface, and multi-surface sag tolerance analysis of optical systems can be achieved by repeatedly applying the algorithm with pre-assigned wave aberration for each surface. Future work will focus on extending the method to compute surface sag tolerance for all optical surfaces simultaneously in a single calculation process, ensuring high computation accuracy and efficiency. Furthermore, critical ray has the potential to be integrated with the differentiable optics\cite{wang2022differentiable}. This integration can resolve critical rays using differential information, further accelerating the critical ray computation process and offering guidance for the desensitization design of optical systems.

\begin{backmatter}
\bmsection{Funding} National Key Research and Development Program of China (2023YFC2414700); National Natural Science Foundation of China (12274156); Science, Technology and Innovation Commission of Shenzhen Municipality (JCYJ20210324115812035); The Special Program of Science and Technology Innovation Talent and Service in Hubei Province (2023EHA028)

\bmsection{Disclosures} The authors declare no conflicts of interest.


\bmsection{Data Availability} Data underlying the results presented in this paper are not publicly available at this time but may be obtained from the authors upon reasonable request.

\end{backmatter}

\bibliography{sample}
\bibliographyfullrefs{sample}



\ifthenelse{\equal{\journalref}{aop}}{%
\section*{Author Biographies}
\begingroup
\setlength\intextsep{0pt}
\begin{minipage}[t][6.3cm][t]{1.0\textwidth} 
  \begin{wrapfigure}{L}{0.25\textwidth}
    \includegraphics[width=0.25\textwidth]{john_smith.eps}
  \end{wrapfigure}
  \noindent
  {\bfseries John Smith} received his BSc (Mathematics) in 2000 from The University of Maryland. His research interests include lasers and optics.
\end{minipage}
\begin{minipage}{1.0\textwidth}
  \begin{wrapfigure}{L}{0.25\textwidth}
    \includegraphics[width=0.25\textwidth]{alice_smith.eps}
  \end{wrapfigure}
  \noindent
  {\bfseries Alice Smith} also received her BSc (Mathematics) in 2000 from The University of Maryland. Her research interests also include lasers and optics.
\end{minipage}
\endgroup
}{}

\end{document}